\documentclass[runningheads]{llncs}
\usepackage[T1]{fontenc}
\usepackage{graphicx}
\usepackage{amsmath,amssymb,mathtools}
\usepackage{siunitx}
\usepackage{microtype}
\usepackage{booktabs}
\usepackage{graphicx}
\usepackage{xspace}
\usepackage{xcolor}
\usepackage{multirow}

\usepackage{algorithm}
\usepackage{algpseudocode} 

\usepackage[section]{placeins}  

\newcommand{\method}{FLAIR}
\newcommand{\archi}{Cluster based Federated Learning}
\newcommand{\CH}{CH}
\newcommand{\VRF}{VRF}
\usepackage{hyperref}
\hypersetup{colorlinks=true,linkcolor=black,citecolor=black,urlcolor=black}

\begin{document}
\title{\LARGE \method : Distributed Federated Learning with Dynamic Clustering}
%
%
\author{Ihssan Boutebicha \and Bilel Zaghdoudi \and Mohamed Amine Legheraba \and Maria Potop-Butucaru}
%

%
\institute{LIP6, Sorbonne Université \\
Paris, Île-de-France, F-75005, France \\
\email{Name.Surname@lip6.fr}}
\titlerunning{FLAIR: Distributed Federated Learning with Dynamic Clustering}
\authorrunning{Boutebicha et al.}
\maketitle              
\begin{abstract}
Federated Learning (FL) offers a privacy-preserving framework for distributed machine learning, yet conventional centralized and hierarchical architectures present significant challenges in terms of scalability, resilience, and single points of failure, particularly in dynamic, infrastructure-less environments such as sensor networks. To address these limitations, we introduce \method, a novel, fully decentralized FL protocol that integrates dynamic, resource-aware, secure, and self organized clustering with in-cluster model training. \method~ uses a probabilistic, verifiable cluster-head election mechanism that is enhanced to promote nodes with greater computational and communication capabilities, ensuring both fairness and efficiency. Through comprehensive simulations in ns-3, we evaluate our method against centralized, hierarchical, and gossip-based FL benchmarks across four demanding scenarios. The results demonstrate the superiority of our approach: in static 100-node networks, \method~ achieves a final accuracy of approximately 0.91, outperforming all baselines. The protocol exhibits exceptional robustness, maintaining graceful degradation with accuracy above 0.85 even under 90\% node failure rates. Furthermore, it shows strong mobility resilience, with a performance loss of less than 2\% compared to static deployments. In a realistic smart farming simulation, \method's accuracy is within 0.2\% of the centralized baseline, confirming its practical viability. These findings validate that \method~ successfully combines the scalability of decentralized learning with the structural efficiency of clustering, presenting a robust and high-performing solution for large-scale, heterogeneous IoT systems.

\end{abstract}
\section{Introduction}~\label{sec:intro}

Federated Learning (FL) has emerged as a powerful paradigm for privacy-preserving machine learning on distributed data~\cite{lim2020fedsurvey}. However, its conventional centralized architecture suffers from critical limitations, including scalability bottlenecks and single points of failure, which motivate decentralized approaches~\cite{kairouz2021fedsurvey,rodriguez2023flattacks}. These challenges are especially pronounced in infrastructure-less environments like wireless sensor networks (WSNs), where node resources and connectivity are dynamic.

In parallel, clustering has been established as a key strategy in WSNs~\cite{swetha2023novel} such as those deployed for environmental monitoring, healthcare (e.g., body area networks), and smart cities, in order to enhance scalability and energy efficiency. Protocols like LEACH~\cite{heinzelman2000leach} pioneered the use of rotating cluster-heads to create an efficient, hierarchical network structure~\cite{heinzelman2000leach,heinzelman2002leachc,younis2004heed,lindsey2002pegasis}. This approach provides a natural foundation for structuring distributed computation and communication.

Combining these two fields offers a promising path for robust decentralized learning. Using dynamic clusters for FL can reduce energy consumption, adapt to network heterogeneity, and mitigate node failures. Yet, despite this potential, many existing methods rely on fixed network structures or external infrastructure, which is inconsistent with the nature of sensor networks. For example, gossip-based aggregation~\cite{ormandi2013gossip}, epidemic-style learning~\cite{devos2024epidemic} or HEAL~\cite{aina2025heal} still assume static overlays or require additional components, but do not fully integrate dynamic clustering. There remains a clear need for a protocol that unifies the principles of FL and dynamic clustering in a fully decentralized and adaptive manner.

To bridge this critical gap, we introduce \method~(Federated Learning with Adaptive Integrity-preserving Randomness), a protocol that advances the state-of-the-art by establishing a synergistic link between secure and self-organized network-layer clustering and application-layer learning. Our design is inspired by the probabilistic cluster-head rotation of the seminal LEACH protocol~\cite{heinzelman2000leach}, which we fundamentally enhance with two key innovations: a resource-aware scoring mechanism and verifiable randomness in the election process. This cross-layer architecture is engineered to deliver a superior balance of energy efficiency, robustness, and learning performance. Our extensive simulations validate this approach, demonstrating that \method~ sustains rapid convergence and high accuracy while significantly enhancing resilience against the node failures and disconnections characteristic of dynamic wireless environments.

\section{Related Work}~\label{sec:related}
The integration of clustering with federated learning has recently emerged as a promising approach for improving communication efficiency and robustness. In clustered FL, clients are organized into groups where model updates are aggregated locally before being propagated to a central coordinator or a higher-level aggregator. Lalitha et al.~\cite{lalitha2019p2pfl} studied peer-to-peer FL on graph topologies, enabling model sharing without centralized coordination. Hierarchical FL~\cite{liu2020hierarchical} utilizes intermediate aggregators (e.g., edge servers or cluster-heads) to improve convergence speed and lower communication overhead. Concurrently, blockchain-based FL~\cite{lu2020blockchainfl} has been proposed as a method for ensuring verifiable trust in multi-party learning settings. Other approaches investigate device-to-device communication and over-the-air aggregation to further reduce latency and bandwidth usage~\cite{xiao2024over}.

Although these studies highlight the advantages of clustering and hierarchical structures in FL, they predominantly rely on static or infrastructure-assisted aggregation models—an assumption that is often impractical in dynamic sensor networks characterized by node mobility and resource variability. To address this limitation, new protocols are needed that combine the principles of dynamic clustering and distributed learning, enabling energy-efficient and adaptive federated learning in wireless environments. Our proposed \method ~addresses these issues. 

This paper makes the following primary contributions to the field:
\begin{itemize}
    \item[\textbullet] A cluster-based FL architecture.
    \item[\textbullet] A novel, decentralized FL protocol, FLAIR, designed for sensor networks that uniquely integrates probabilistic, resource-aware, and verifiable mechanisms for dynamic cluster-head election. This design ensures both fair load distribution and directs aggregation tasks to more capable nodes.
    \item[\textbullet] A comprehensive validation of \method~ conducted in the ns-3 simulator. We rigorously benchmark our protocol against four state-of-the-art baselines—centralized FL~\cite{mcmahan2017communication}, GAIA~\cite{hsieh2017gaia}, HEAL~\cite{aina2025heal}, and Gossip Learning~\cite{ormandi2013gossip}, demonstrating its superior performance across multiple metrics including accuracy, convergence speed, and fault tolerance.
\end{itemize}

The remainder of this paper is organized as follows. Section III introduces our \method~ protocol. Section IV describes the experimental setup and reports the evaluation results. Section V concludes the paper and discusses future research directions.

\section{Proposed \archi~System}~\label{sec:leachfl}
\vspace{-7mm}
\subsection{System overview}

We propose the \archi~ system, a layered architecture for energy-efficient and adaptive federated learning in sensor networks. The system is designed to achieve three primary objectives: (i) robustness, by avoiding single points of failure, (ii) performance, by leveraging nodes with greater computational and communication resources, and (iii) energy efficiency, by restricting transmissions to short-range intra-cluster communication and rotating aggregator roles.

\begin{figure}
    \centering
    \includegraphics[width=0.4\linewidth]{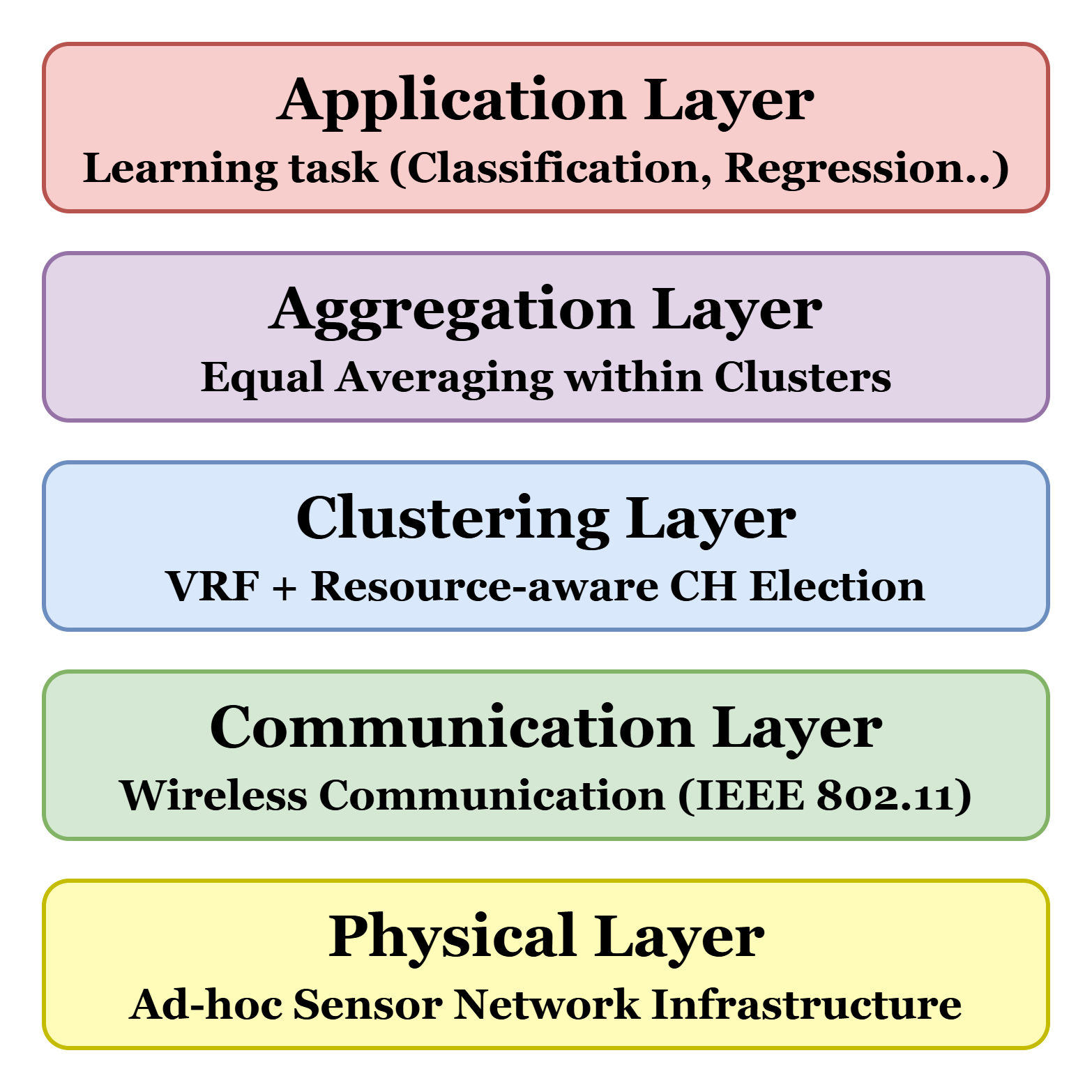}
    \caption{Layered architecture of our \archi}
    \label{fig:architecture}
\end{figure}

Our \archi ~system integrates networking and learning into a unified, five-layer architecture (Figure~\ref{fig:architecture}). The system is built on a Physical Layer of heterogeneous sensor nodes and a Communication Layer for reliable packet delivery (e.g., IEEE 802.11). Above this, the Clustering Layer uses the \method~ protocol to organize nodes into dynamic clusters, where the Aggregation Layer coordinates local model updates. Finally, the top-level Learning Task Layer executes the FL application, such as classification or regression.


Cross-layer interactions are critical: clustering decisions depend on resource information from lower layers, while learning efficiency influences communication and energy usage.

At the core of \archi's Clustering layer lies our proposed \method~ protocol. It is based on the Low-Energy Adaptive Clustering Hierarchy protocol (LEACH)~\cite{heinzelman2000leach} but includes several key extensions: (i) resource-aware CH election, so that more capable devices handle heavier aggregation duties; (ii) verifiable randomness to prevent biased CH election and ensure auditability; and (iii) an in-cluster federated learning workflow. This design integrates network-layer clustering with application-layer learning, forming a unified, layered architecture.

Building on these principles, \method~ operates in rounds, each consisting of two phases: (i) \emph{network clustering}, where nodes self-organize into clusters, and (ii) \emph{federated learning}, where local updates are trained and aggregated within clusters. At the beginning of each round, clusters are reconfigured to adapt to resource variability and mobility, making the protocol suitable for dynamic sensor networks where resource-constrained nodes collaborate without fixed infrastructure.

\subsection{System Model and Assumptions}
We consider a set $\mathcal{N}$ of nodes communicating over a shared wireless medium in an infrastructure-less ad-hoc environment. Each node is heterogeneous in its computational (CPU, GPU, memory) and communication (bandwidth) capabilities, and holds a \emph{private} dataset.\\

Time progresses in rounds $r=1,2,\dots$. In each round, a subset of nodes acts as cluster-heads, while the remaining nodes join exactly one cluster. Communication is single-hop within clusters (i.e., each member communicates directly with its \CH, and \CH s broadcast directly to their members). No inter-cluster coordination is required, which reduces protocol overhead and eliminates additional failure points.

Two mechanisms are central to the protocol. First, a target fraction $p \in (0,1)$ of nodes is expected to serve as \CH s in each round, ensuring probabilistic load balancing across the network. Second, every node can evaluate a Verifiable Random Function (\VRF), which produces a publicly verifiable random value in $[0,1]$. This primitive prevents manipulation of elections and guarantees fairness, while the process remains lightweight enough for resource-constrained environments.

Together, these assumptions provide the foundation on which \method ~integrates clustering and federated learning: clustering organizes the communication structure efficiently, while FL ensures collaborative training without exposing raw data.

\subsection{Phase 1: Network Clustering}


\subsubsection{Step 1: Cluster-Head Selection}
Let $G$ be the set of eligible nodes at round $r$. A node $n$ is eligible to become a \CH ~only if it has not acted as one during the last $\frac{1}{p}$ rounds. Each node in $G$ computes a threshold $T(n)$ and samples $x \sim \mathrm{Uniform}(0,1)$ through a \VRF. Node $n$ elects itself as \CH if $x < T(n)$. The threshold incorporates a \emph{resource score} $R_n$, which biases elections toward resource-rich nodes:

\vspace{1mm}
\begin{equation}
\label{eq:Tn}
T(n) =
\begin{cases}
\dfrac{p \cdot R_n}{1 - p \cdot \left(r \bmod \frac{1}{p}\right)} & \text{if } n \in G,\\[1.25ex]
0 & \text{otherwise.}
\end{cases}
\end{equation}
where
\begin{equation}
\label{eq:Rn}
\begin{aligned}
R_n &= \alpha \cdot CPU_n + \beta \cdot RAM_n + \gamma \cdot GPU_n + \delta \cdot BW_n, \\[1ex]
&\text{With } \alpha + \beta + \gamma + \delta = 1.
\end{aligned}
\end{equation}
\vspace{0.5mm}

The variables $CPU_n, RAM_n, GPU_n,$ and $BW_n$ are normalized resource indicators. As summarized in Algorithm~\ref{alg:ch-selection}, this weighted, probabilistic scheme favors more capable nodes for CH duty while preserving the load balancing of LEACH~\cite{heinzelman2000leach}. The use of a \VRF ensures the election is fair and tamper-resistant.

\begin{algorithm}[tbp]
\caption{Resource-Aware \CH{} Selection with Verifiable Randomness}
\label{alg:ch-selection}
\begin{algorithmic}[1]
\Require target \CH{} ratio $p \in (0,1)$; round $r$; eligibility set $G$; resource vector $(CPU_n,RAM_n,GPU_n,BW_n)$; weights $\alpha,\beta,\gamma,\delta \ge 0$
\For{each node $n \in \mathcal{N}$ \textbf{in parallel}}
  \If{$n \notin G$}
    \State $T(n) \gets 0$
  \Else
    \State $R_n \gets \alpha\cdot CPU_n + \beta\cdot RAM_n + \gamma\cdot GPU_n + \delta\cdot BW_n$
    \State $T(n) \gets \dfrac{p \cdot R_n}{1 - p \cdot (r \bmod \frac{1}{p})}$
  \EndIf
  \State $x \gets \VRF\_Uniform(0,1)$ \Comment{publicly verifiable}
  \If{$x < T(n)$}
    \State $n$ broadcasts \textsf{CH-ADV} (cluster-head advertisement)
    \State mark $n$ as \CH{}
  \EndIf
\EndFor
\end{algorithmic}
\end{algorithm}

\subsubsection{Step 2: Cluster Formation}
Elected \CH s broadcast advertisements (\textsf{CH-ADV}). As shown in Algorithm~\ref{alg:cluster-formation}, non-\CH~ nodes join the most suitable \CH~ based on a distance-based cost, forming stable clusters with balanced resource allocation.
\begin{algorithm}[tbp]
\caption{Cluster Formation}
\label{alg:cluster-formation}
\begin{algorithmic}[1]
\For{each non-\CH{} node $u$ \textbf{in parallel}}
  \State listen for \textsf{CH-ADV} beacons; collect candidate set $\mathcal{C}$
  \State compute $cost(u,c)$ for all $c\in\mathcal{C}$ 
  \State $c^\star \gets \arg\min_{c\in\mathcal{C}} cost(u,c)$
  \State send \textsf{JOIN-REQ} to $c^\star$
\EndFor
\For{each \CH{} $c$}
  \For{each \textsf{JOIN-REQ} from $u$}
     \State admit $u$ if capacity allows; send \textsf{JOIN-ACK}
     \State update membership list 
  \EndFor
\EndFor
\end{algorithmic}
\end{algorithm}
\begin{algorithm}[tbp]
\caption{In-Cluster Federated Learning }
\label{alg:fl}
\begin{algorithmic}[1]
\Require initial model $\mathbf{w}^{(0)}$, batch size $B$, in-cluster epochs $E_\text{round}$
\For{each cluster $c$ in parallel}
  \State $t \gets 0$
  \For{$e = 1$ to $E_\text{round}$}
    \For{each member $u \in \mathcal{S}_c$ \textbf{in parallel}}
       \State $\mathbf{w}_u^{(t+1)} \gets \textsf{LocalTrain}(\mathbf{w}^{(t)}, \mathcal{D}_u, B)$
       \State send $\mathbf{w}_u^{(t+1)}$ to \CH
    \EndFor
    \State \CH~ computes $\mathbf{w}^{(t+1)}$ via \eqref{eq:simpleavg}
    \State \CH~ broadcasts $\mathbf{w}^{(t+1)}$ to $\mathcal{S}_c$
    \State $t \gets t+1$
  \EndFor
\EndFor
\end{algorithmic}
\end{algorithm}
\subsection{Phase 2: Federated Learning}
Once clusters are formed, each cluster runs an independent learning process, removing the need for global coordination.

\subsubsection{Step 1: Local Training}
Every node $u$ trains the current model $\mathbf{w}^{(t)}$ on its private dataset $\mathcal{D}_u$ for $E$ local epochs, producing a locally updated model $\mathbf{w}_u^{(t+1)}$. Only updates are shared; raw data always remains private.

\subsubsection{Step 2: Model Aggregation}
Each \CH~ aggregates updates from its members using a simple averaging rule, as shown in Equation~\ref{eq:simpleavg}. The aggregated model is then redistributed to all cluster members for the next iteration.
\begin{equation}
\label{eq:simpleavg}
\mathbf{w}^{(t+1)} = \frac{1}{|\mathcal{S}_c|} \sum_{u \in \mathcal{S}_c} \mathbf{w}_u^{(t+1)},
\end{equation}

\subsubsection{Step 3: Iterative In-Cluster Training}
The local training and aggregation steps are repeated for a fixed number of epochs ($E_\text{round}$) per round. This iterative in-cluster process, detailed in Algorithm~\ref{alg:fl}, constitutes one full round of federated learning.

\section{Experimental Evaluation}~\label{sec:eval}
\vspace{-10mm}
\subsection{Implementation Setup}
Our \method~ protocol was implemented in the \texttt{ns-3} simulator, with all baselines re-implemented in the same environment to ensure strict comparability. Experiments were executed on a dedicated server equipped with two Intel Xeon E5-2660 v3 processors, each providing 10 cores and 2 threads per core for a total of 40 logical processors operating at a base frequency of 2.6\,GHz and a maximum turbo frequency of 3.3\,GHz. The system had 125\,GB of RAM and a 456\,GB main storage partition, running Arch Linux (kernel 6.12.4-arch1-1). 
\begin{table*}[h]
\centering
\makebox[\textwidth][c]{%
\begin{tabular}{p{2cm} p{6.3cm} p{7cm}}
\toprule
\textbf{Test} & \textbf{Objective} & \textbf{Setup} \\
\midrule
Experiment~1 & Comparative evaluation in static networks & 100 static nodes; comparison with C-FL, GAIA, HEAL, and Gossip Learning. \\
Experiment~2 & Resilience to node dropouts & Permanent, temporary, and random crashes (up to 90\% of nodes); two settings ($E=1$, $E=3$). \\
Experiment~3 & Impact of mobility on learning performance & 5 mobility models; perfect vs. range-limited connectivity. \\
Experiment~4 & Smart farming with heterogeneous nodes & 80 fixed sensors + 20 mobile robots; Kaggle ``Predicting Watering the Plants'' dataset. \\
\bottomrule
\end{tabular}}
\caption{Overview of the experimental tests conducted for evaluating \method~}
\label{tab:tests-overview}
\end{table*}

\vspace{-5mm}

We used two datasets: the ``\textit{Spambase} dataset''~\cite{hopkins1999spambase}, a standard binary classification benchmark, and the ``\textit{Predicting Watering the Plants}'' dataset from Kaggle~\cite{nelakurthi2021plants} for a smart farming scenario. The simulated network consisted of $100$ heterogeneous wireless nodes using IEEE~802.11 in ad-hoc mode. A logistic regression model was the learning objective~\cite{hosmer2013applied}. For the resource score (Equation~\ref{eq:Rn}), the four weights were set to $0.25$, giving equal importance to CPU, RAM, GPU, and bandwidth. We designed four tests to assess performance, summarized in Table~\ref{tab:tests-overview}. 
To account for stochastic variability inherent to wireless simulation, each configuration was executed multiple times with different random seeds, and the reported curves correspond to the empirical mean across these independent runs.

\subsection{Comparative Evaluation in Static Networks}
To evaluate the performance of \method ~, we compared it against four representative distributed protocols: Centralized FL (C-FL)~\cite{mcmahan2017communication}, the canonical client-server paradigm; GAIA~\cite{hsieh2017gaia}, a hierarchical architecture; HEAL~\cite{aina2025heal}, a hybrid approach with dynamically elected hubs; and Gossip Learning~\cite{ormandi2013gossip}, a fully decentralized peer-to-peer protocol. All baselines were carefully tuned for competitive performance, and their configurations are summarized in Table~\ref{tab:baseline-params}.

\begin{table}[tbp]
\centering
\caption{Configuration of baseline protocols in Experiment~1}
\label{tab:baseline-params}
\begin{tabular}{lll}
\toprule
\textbf{Protocol} & \textbf{Aggregation Method} & \textbf{Parameters} \\
\midrule
C-FL      & Global Avg           & $E=3$, $\eta=0.01$ \\
GAIA      & Local + Global Avg   & $5$ local servers, $E=3$ \\
HEAL      & Hub + Inter-hub      & $5$ hubs/round \\
Gossip    & Pairwise Averaging   & $3$ peers/round \\
\method ~& In-cluster Averaging & $E=3$ \\
\bottomrule
\end{tabular}
\end{table}

\begin{figure}[tbp]
    \centering
    \includegraphics[width=.6\linewidth]{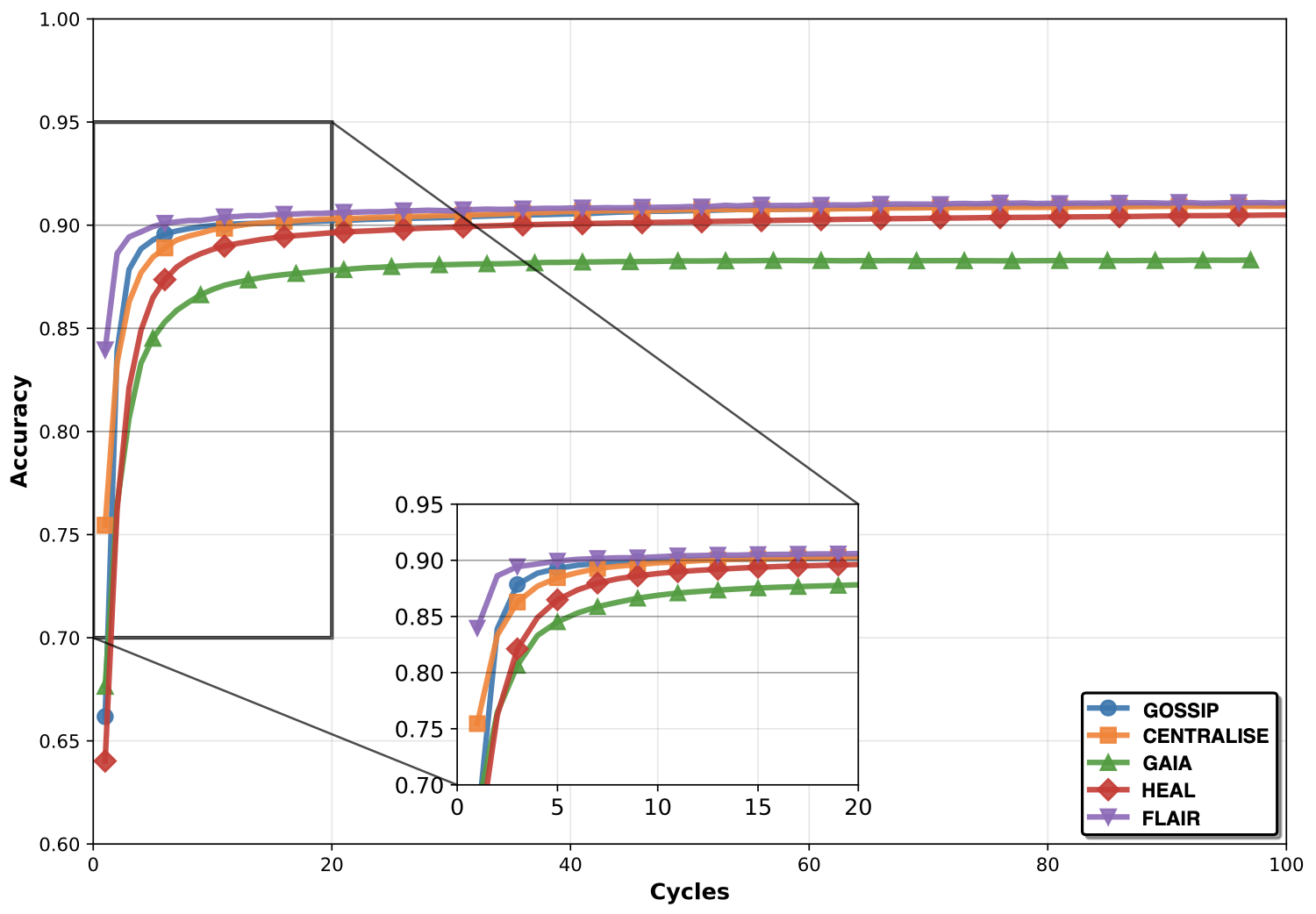}
    \caption{Accuracy evolution of \method ~and baselines in static networks (100 nodes)}
    \label{fig:test1}
\end{figure}

Figure~\ref{fig:test1} shows the accuracy evolution over training cycles in a static network of $100$ nodes. The inset highlights early convergence dynamics. \method ~achieves the highest final accuracy ($\sim 0.91$), surpassing C-FL, HEAL, and Gossip ($\sim 0.90$), and clearly outperforming GAIA ($\sim 0.88$).

These results demonstrate that the clustering-based design of \method ~accelerates convergence while sustaining higher steady-state accuracy. Compared to GAIA, convergence is up to $2.5\times$ faster, and compared to Gossip Learning, the protocol requires significantly fewer cycles to stabilize. Overall, \method ~combines the scalability of decentralized designs with the efficiency of clustering, providing superior performance in static deployments.

\subsection{Resilience to Node Dropouts}

\begin{figure*}
    \centering
    \makebox[\textwidth][c]{\includegraphics[width=1.3\textwidth]{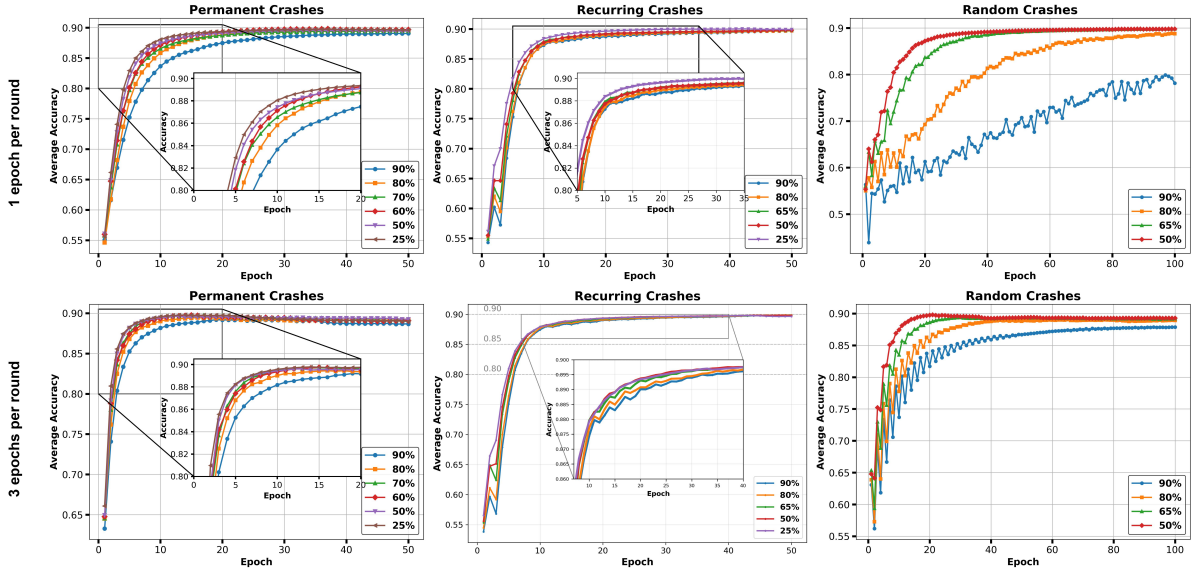}}
    \caption{Accuracy evolution of \method ~under different node dropout conditions}
    \label{fig:test2}
\end{figure*}

We next examined the resilience of \method ~to node failures by simulating three dropout types: permanent crashes, temporary crashes, and random dropouts. Performance was assessed based on final accuracy, recovery speed, and stability. To capture the effect of round duration, we considered both long ($E=3$) and short ($E=1$) round configurations. Figure~\ref{fig:test2} shows the results.

\begin{table}
    \centering
    \caption{Number of rounds required to reach 0.80 and 0.85 accuracy under different dropout scenarios and round durations}
    \label{tab:test2}
    \begin{tabular}{lcccc}
        \toprule
        \multirow{2}{*}{\textbf{Scenario}} & \multicolumn{2}{c}{\textbf{$E=3$}} & \multicolumn{2}{c}{\textbf{$E=1$}} \\
        \cmidrule(lr){2-3} \cmidrule(lr){4-5}
         & \textbf{0.80} & \textbf{0.85} & \textbf{0.80} & \textbf{0.85} \\
        \midrule
        Permanent crashes (80\%) & 3 & 4 & 5 & 9 \\
        Temporary crashes (80\%) & 4 & 8 & 6 & 9 \\
        Random crashes (80\%) & 4 & 10 & 39 & 55 \\
        Permanent crashes (90\%) & 4 & 6 & 8 & 13 \\
        Temporary crashes (90\%) & 5 & 8 & 7 & 10 \\
        Random crashes (90\%) & 12 & 29 & -- & -- \\
        \bottomrule
    \end{tabular}
\end{table}
\vspace{-10mm}
In the baseline case without dropout, \method ~stabilized at around $0.90$ accuracy. The protocol showed graceful degradation under permanent crashes, converging above $0.88$ even with $90\%$ node loss, as detailed in Table~\ref{tab:test2}. While temporary crashes caused only minor, recoverable perturbations, random crashes were more disruptive, especially with short rounds ($E=1$). In contrast, longer rounds ($E=3$) smoothed the learning dynamics and improved stability because the longer aggregation intervals absorbed much of the instability. Overall, \method ~consistently maintained accuracy above $0.85$ across all scenarios, confirming its strong resilience even under extreme dropout conditions.
\subsection{Impact of Mobility on Learning Performance}

\begin{figure}
    \centering
    \includegraphics[width=.6\linewidth]{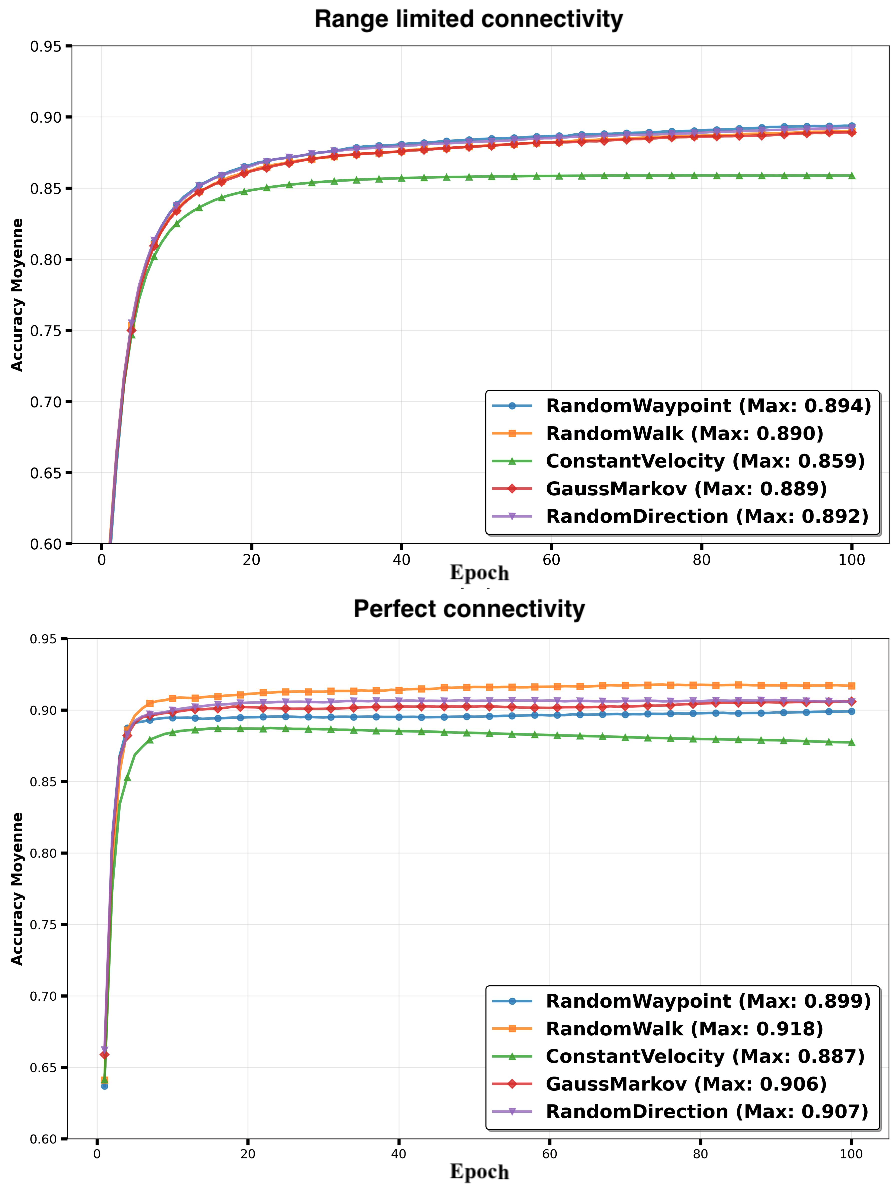}
    \caption{Accuracy evolution of \method ~under five mobility patterns, comparing perfect and range-limited connectivity}
    \label{fig:test3}
\end{figure}

We evaluated the effect of node mobility on \method~ using five standard mobility models~\cite{bai2004survey} under two scenarios: perfect and range-limited connectivity. As shown in Figure~\ref{fig:test3}, mobility had no measurable impact with perfect connectivity. With range-limited connectivity, occasional disconnections caused only minor perturbations, with overall accuracy remaining within $2\%$ of the static baseline (staying above $0.88$ for all patterns). These results indicate that \method~ is resilient to mobility and that its clustering mechanism adapts effectively to dynamic topologies.

\subsection{Smart Farming with Heterogeneous Nodes}
\begin{figure}
    \centering
    \includegraphics[width=.6\linewidth]{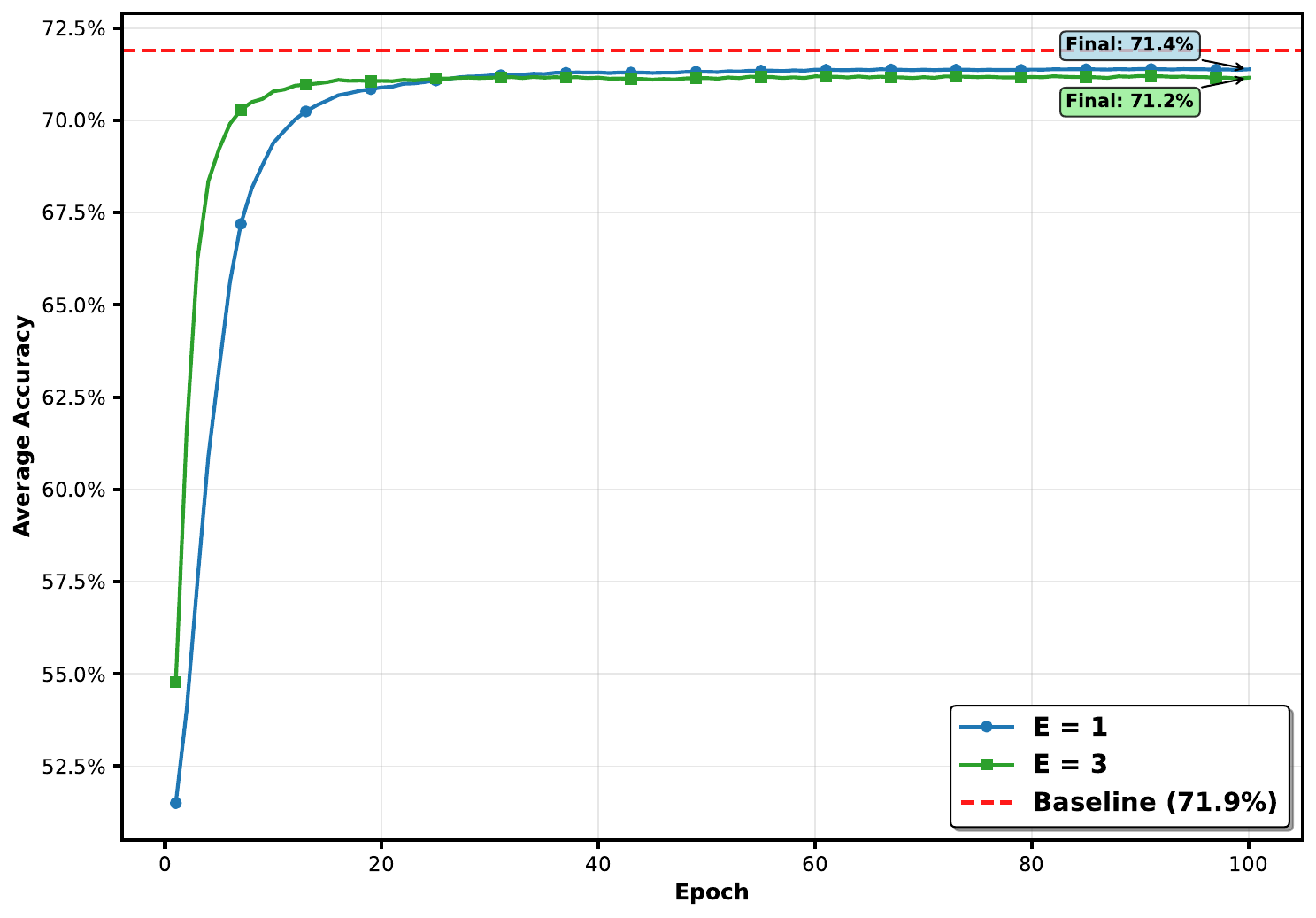}
    \caption{Accuracy evolution on the "Predicting Watering the Plants" dataset, under the smart farming setup (80 fixed sensors + 20 mobile robots). Two local update settings ($E=1$ vs. $E=3$) are compared against the centralized baseline (71.9\%).}
    \label{fig:test4}
\end{figure}

\begin{figure*}[]
    \centering
    \makebox[\textwidth][c]{\includegraphics[width=1.3\textwidth]{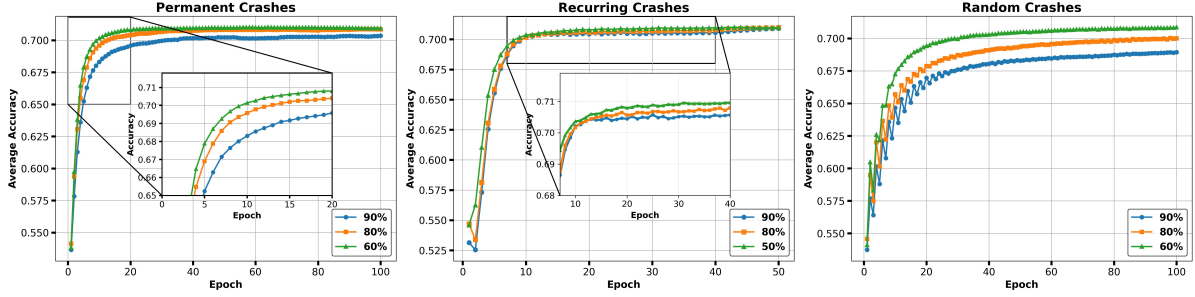}}
    \caption{Accuracy evolution of \method ~under different node dropout conditions in the smart farming scenario}
    \label{fig:test4.2}
\end{figure*}

To evaluate \method ~in a realistic application, we simulated a smart farm with 80 fixed sensors and 20 mobile robots, using the ``Predicting Watering the Plants'' dataset from Kaggle~\cite{nelakurthi2021plants}. As shown in Figure~\ref{fig:test4}, both long-round ($E=3$) and short-round ($E=1$) configurations converge near the centralized baseline (71.9\%), achieving final accuracies of 71.4\% and 71.2\%, respectively. We also extended the dropout experiments to this setup, and the results (Figure~\ref{fig:test4.2}) confirm that the resilience properties observed in static networks extend to this heterogeneous, application-driven scenario. \method ~demonstrates graceful degradation and rapid recovery even with mobility and partial connectivity, highlighting its practical applicability and robust performance in dynamic IoT environments.

\section{Conclusion}~\label{sec:conclusion}

In this paper, we introduced \method ~, a decentralized federated learning protocol designed to operate efficiently in resource-constrained, wireless networks. By combining principles from network-layer clustering and application-layer distributed learning, \method ~overcomes the inherent weaknesses of centralized FL architectures, such as scalability bottlenecks and single points of failure. The protocol's core innovation lies in its dynamic, resource-aware, and verifiable cluster-head election mechanism, which facilitates load balancing and enhances system robustness without requiring fixed infrastructure.

Our comprehensive experimental evaluation, conducted within the ns-3 simulation environment, has validated the effectiveness of \method ~. Across a series of rigorous experiments, our protocol consistently demonstrated superior performance. In static networks, it achieved faster convergence and higher final accuracy than canonical centralized, hierarchical, and gossip-based FL approaches. Crucially, \method ~exhibited remarkable resilience, sustaining high learning accuracy under extreme node dropout scenarios—up to 90\% failures—and adapting seamlessly to various mobility models with minimal performance degradation. Its successful application to a realistic smart farming scenario further underscored its practical utility in heterogeneous IoT deployments, where it achieved performance nearly identical to a centralized baseline.

\section{Acknowledgments}
This work was funded by the European Union’s Horizon Europe project CyberNEMO (101168182).
\bibliographystyle{splncs04}
\bibliography{mybibliography}
\end{document}